\documentclass[conference]{IEEEtran}

\usepackage{enumerate}
\usepackage{algorithm}
\usepackage[noend]{algpseudocode}
\usepackage{algorithmicx}

\algblock{ParFor}{EndParFor}
\algnewcommand\algorithmicparfor{\textbf{parfor}}
\algnewcommand\algorithmicpardo{\textbf{do}}
\algnewcommand\algorithmicendparfor{\textbf{end\ parfor}}
\algrenewtext{ParFor}[1]{\algorithmicparfor\ #1\ \algorithmicpardo}
\algrenewtext{EndParFor}{\algorithmicendparfor}
\usepackage[font=small, labelfont=bf]{caption}

\usepackage{cite}
\usepackage[pdftex]{graphicx}
 
\usepackage{pgfplots}
\usepackage{times}
\usepackage{listings}
\usepackage{float}
\begin{document}

\lstset{language=C++,basicstyle={\small\ttfamily}, linewidth= 8.7cm}

%\title{Analyzing Parallel Programming Models for Analysis of Small Object Orbital Data}
%\title{Nested Load Imbalance: Comparison of Strategies and Programming Languages for Parellelization of an Orbital Analysis Code}
%\title{Coordinating Nested Parallelism and Extreme Load Imbalance in an Orbital Analysis Code}
\title{Handling Nested Parallelism and Extreme Load Imbalance in an Orbital Analysis Code}

%% author names and affiliations
%% use a multiple column layout for up to three different
%% affiliations
\author{
\IEEEauthorblockA{Benjamin James Gaska, Neha Jothi, Mahdi Soltan Mohammadi \\and Michelle Mills Strout}
\IEEEauthorblockA{Computer Science\\
University of Arizona\\
Tucson, Arizona\\
Email: \{bengaska, nehaj, mahdi.s.m.k, mstrout\}@email.arizona.edu}
\and
\IEEEauthorblockN{Kat Volk}
\IEEEauthorblockN{Lunar and Planetary Laboratory\\
University of Arizona\\
Tucson, Arizona\\
Email: kvolk@lpl.arizona.edu}
}

\maketitle

%%%%%%%%%%%%%%%%%%%%%%%%%%
\begin{abstract}
Nested parallelism exists in scientific codes that are searching multi-dimensional
spaces.  However, implementations of nested parallelism often have 
overhead and load balance issues.  The Orbital Analysis code we present
exhibits a sparse search space, significant
load imbalances, and stopping when the first solution is reached.
All these aspects of the algorithm
exacerbate the problem of using nested parallelism effectively.
In this paper, we present an inspector/executor strategy for 
chunking such computations into parallel wavefronts.
The presented shared memory parallelization is no longer nested and
exhibits significantly less load imbalance.
We evaluate this approach on an Orbital analysis code,
and we improve the execution time from the original implementation by
an order of magnitude.
As part of a Graduate Computer Science course in Parallel Programming models,
we show how the approach can be implemented in 
parallel Perl, Python, Chapel, Pthreads, and OpenMP.
Future work includes investigating how to automate and generalize
the parallelization approach.

% In this paper, we parallelize the
%computation using various parallel programming models and are able
%to improve
%the execution time of the analysis by more than an order of magnitude
%thus enabling the analysis that will be run once a month to execute in
%around one day. We identify key problems in this workload in terms of
%parallelization and experiment with some possible methods that help with
%the significant load imbalance. We show how the parallelization with
%more regular load balance can be implemented in
%parallel Perl, Python, Chapel, Pthreads, and OpenMP.
%This case study provides information to aid in the selection of the
%programming model for the production code.
\end{abstract}

\IEEEpeerreviewmaketitle

%%%%%%%%%%%%%%%%%%%%%%%%%%%
%\section{Questions}
%
%\begin{itemize}
%\item Who else is writing analysis codes for discovering orbital resonance and libration and
%what programming language are they using?
%People describe what analyses they are doing but not how they are coding it up and 
%software not typically shared. Publish results of such analyses.
%Typically Fortran is what is posted.
%Observational survey posting Fortran and Python code to github.
%Saw these objects, what are they doing?  Code for classifying might not be publically
%available due to support reasons.
%
%\item How much variance is there between objects that take require checking most of the
%ratios and those that require few checks?  What are some concrete numbers?
%Kat might be able to give us a distribution on this. (60\% ish)
%
%\item Can put into models about history of Neptune.
%
%\item What does it mean when we say that the classification has happened manually?
%Every single classification is still checked by an eye.  Could do a second check?
%
%\item Year that LSST should come online?  5 to 10 years from now.
%
%\item Do we only need to do the analysis for each object once?
%Everytime fit the orbit with new observations.  Whole sky once per month.
%\end{itemize}

%%%%%%%%%%%%%%%%%%%%%%%%%%
\section{Introduction}

Nested loop parallelism is natural to express in programming models such
as OpenMP, but difficult to efficiently realize when sparse computation
spaces with significant load imbalances and early termination criteria are involved.
In this paper we present an approach to parallelizing such computations
on shared memory machines.

%%%%%%%
\begin{figure}
%\centering
\lstset{morekeywords={parfor},language=C,numbers=left,numberstyle=\small,xleftmargin=5.0ex}
\begin{lstlisting}[frame=single]
Main()
  for each part in particle
   if( isConsistent(part) )
     result = CheckResonance(part)
EndMain

checkResonance(part)
 for(p=1;p<pmax;p++)
  for(q=p;p-q<pmax && q>0;q--)
   if( ratio(p,q) not in ratios )
    for(m=q;p-q>=0;m--)
     for(n=m;p-q-m>=0;n--)
      for(r=n;p-q-m-n>=0;r--) {
       s=p-q-m-n-r
       if(checkLibration(part,p,q,m,n,r,s))
        return (p,q,m,n,r,s)
       else
        continue
      }
\end{lstlisting}
\caption{Pseudocode showing the iteration space. Each
particle is evaluated separately. Inside the checkResonance function
we see the deeply nested structure that where the vast majority
of run time occurs.}
\label{orbitOverview}
\end{figure}
%%%%%%%

The orbital analysis code we worked with in this case study 
consists of a six-deep nested loop structure including the outermost
loop over particles (see Fig.~\ref{orbitOverview}).
Each of the six nested loops can be executed in parallel, thus
the computation experiences significant nested parallelism.
One problem is that the iteration space is sparse: particles
are checked for consistency and the {\tt (p,q)} ratio is checked to avoid
equivalent repeats.
Specifically, there is a condition checked at line 10 before the computation
for a particular {\tt (p,q)} iteration executes.
Most of the points in the parameter space fail this check.  Another
problem is early termination:
this code will return upon finding parameters that satisfy the check in
the innermost loop at line 16.

One parallelization alternative is to compute 
each particle independently of the others.
In other words, parallelize the outermost loop and leave all else unchanged.
However, this performs poorly due to load balancing issues.
In the most extreme cases, a single particle can run 2-3 orders of
magnitude longer than the a short running particle.
The execution time for a given particle cannot be predicted without running
it fully.
In the worse case, in which a single processor is given all of the 
long running particles, no gains are achieved from
parallelization.

%%% MMS: bit of a tangent.
%Parallelism was then moved into the nested structure in the checkResonance()
%function, and analysis was done to determine the best depth at which to
%perform our parallelism.
%For the example in Figure~\ref{orbitOverview} the best nesting
%depth for this was the {\tt m} loop.  Therefore, the {\tt p, and q} loops
%remain serial and the set of inner loop parameters that will be used in
%a checkLibration call will be executed with one level of parallelism.
%The reduction step determines the lexicographically minimum set of
%parameters that satisfies the condition if any do in that wavefront.
%This minimum set represents the optimum solution for a given particle.

A second alternative would be to use nested parallelism.
However, the early termination check in the innermost loop makes nested
parallelism impractical.  If all size loops where specified as parallel,
all iterations would execute even when early termination is possible,
which is frequent.
Additionally, there would need to be a reduction computation that determines
the earliest values of {\tt (p,q,m,n,r)} where the condition was satisfied,
because that is the correct result for the program.

A third alternative is to use  task-based parallelism.
This would work by spawning off tasks for each
call to checkLibration(). While implementing the task-based model two problems arise.
First, the non-determinism introduced by the task parallelism
loses the ordering information guaranteed by the serial code (i.e., early termination strikes again). 
This prevents us from knowing if the first value returned is the optimal solution.
Second, there is too much task overhead.
Each call to checkLibration is lightweight, but the amount of calls made is high.
In an extreme case one particle spawned over 140,000 tasks. 
%This led to communication costs overwhelming
%possible gains in the execution time. 
%Therefore, both the nested parallelism and 
%task-based models were deemed non-viable.

To handle the load imbalance, sparse iteration space, and early termination
issues, we developed a finer-grained parallelism
internal to each particle.
The approach consists of building parallel 
wavefronts of tuples in the search space at a particular nesting depth.
Figure~\ref{alg:InternalPseudo} shows the pseudocode for the algorithm implemented.
In the new algorithm, for each {\tt (p,q)} ratio that passes the
check on Line 10, a subset of the search space is collected.
The {\tt CheckLibration()} function can then be called on that 
subset in parallel.  The final loop at Line 23 will check if any of the
tuples in the just executed wavefront passed the libration check and 
thus the computation should  terminate early.

The original orbital analysis code was implemented in Perl and
would have taken more than a month to analyze the 
monthly observations from a new telescope.
Parallelizing the computation and porting the Perl analysis script
to more efficient programming models
makes the execution time practical (less than a week).
%In this paper, we identify the irregular workload associated 
%with each object and the 
%extreme variance that cannot be identified until execution time.
As part of a Graduate Parallel Programming Models class,
we evaluated the process of implementing the parallel analysis script
in Perl, Python, Chapel, Pthreads, and OpenMP.
We compare  code snippets from the various
programming models to exhibit different approaches
for implementing the
presented parallelization strategy

%The ports were done in the context of a graduate course studying
%parallel programming models.  We evaluate the ported programs
%from the perspective of the programming language learning curve,
%how realistic the language is for planetary scientists,
%the ease of use of the parallel constructs, and the performance.

Significant speed-up was achieved over the original Perl version.
Using a 12-Core machine over 3x speed-up was achieved in every implementation
versus their own serial version.
The PThreads version was the overall fastest, bringing the 
worst case per particle down from 541 second to 5.2 seconds, 
a 4x speed-up versus its own serial version, and 103x speed-up versus the 
original Perl baseline.
This improvement allows for the total analyses of a month's worth of data 
(approximately 40,000 particles) to be performed in several hours.

The Astronomy community 
%is not likely to start 
%programming using Pthreads, but the community 
has been developing more software in Python.
The serial Python version performs comparably to the baseline Perl.
With the described algorithm the total execution time for the most
costly particle was still brought down
to 78.8 seconds, or about a 6.5x speedup over the original Perl code.
This brings the computation down into the realm of acceptability.

%%%%%%%%%%
\begin{figure}[t]
\lstset{morekeywords={parfor},language=C,numbers=left,numberstyle=\small,xleftmargin=5.0ex}
\begin{lstlisting}[frame=single]
Main()
  for each part in particle
   if( isConsistent(part) )
     result = CheckResonance(part)
EndMain

checkResonance(part){
 for(p=1;p<=pmax;p++)
  for(q=p; p-q<pmax && q>0; q--)
   if( ratio(p,q) not in ratios ) {
     subset = []
     for(m=p-q; p-q>=0;m--)
      for(n=p-q-m; n >= 0; n--)
       for(r=p-q-m-n; r >= 0; r--) {
         s = p-q-m-n-r
         subset.append((p,q,m,n,r,s))
        }
     
     // parallel wavefront
     posSols = []
     parfor(i=0;i<len(subset);i++)
      possSols[i]=CheckLibration(subset[i])
     for(i=0;i<len(possSols);i++)
      if( possSols[i] )
       return subset[i]
   }
EndCheckResonance
\end{lstlisting}
\caption{Pseudocode for building a list of values and searching over the space. This allows us to factor the 
return statement out of the loops, allowing for parallelization to be applied on the internal structure.}
\label{alg:InternalPseudo}
\end{figure}
%%%%%%%%%%

In this paper, we make the following contributions:

%\begin{enumerate}[I]
 %  \setcounter{enumi}{1}
   \begin{itemize}
    \item Provide a description of the application: details about the
          problems faced by the scientist that necessitate code performance improvement.
    \item Parallelizing the code: Why simple solutions do not work, discussion
          of the inherent workload issues, and the final solution.
    \item Description of how the parallelization approach can be mapped to the 
    programming constructs in various programming languages.
    \item Comparison of implementations across different parallel programming models
    in terms of performs.
\end{itemize}

\begin{figure*}[t] %[htbp]
   \centering
   \includegraphics[width=6.2in]{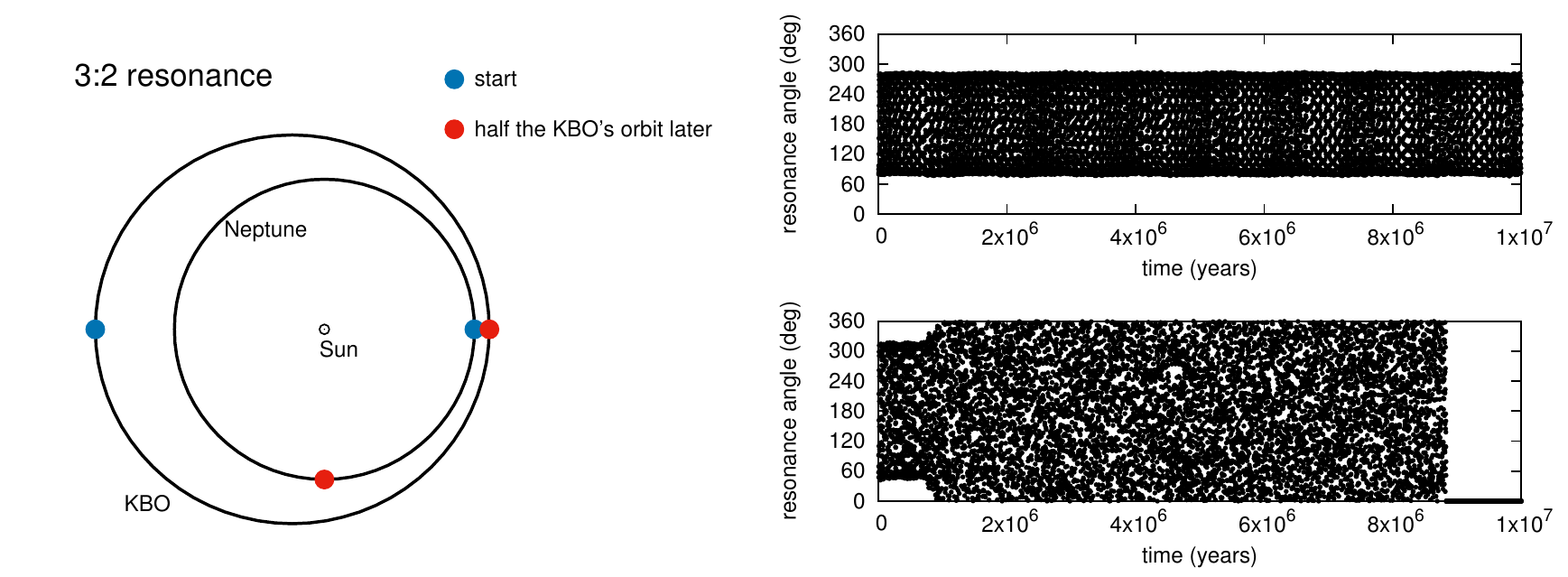}
	\caption{Left: An illustration of the geometry of a 3:2 orbital resonance 
	with Neptune. The Kuiper belt object's (KBO) orbital period is 1.5 times 
	Neptune's orbital period; this means that while the KBO completes half 
	an orbit, Neptune completes 3/4 of an orbit. Right: The resonance angle 
	for an object in the 3:2 resonance (top) and one that starts in the resonance, 
	but does not stay in resonance (bottom). }
   \label{f:res-example}
\end{figure*}

%%%%%%%%%%%%%%%%%%%%%%%%%%
\section{Orbital Analysis Application Details}

%Updated by Dr. Volk
The Kuiper belt is a population of small bodies in the outer solar system.
When new objects are discovered in the Kuiper belt, a common first
step is to simulate the orbits of the observed objects forward in time
and then analyze the results to distinguish between various types of
orbital evolution.  The problem is that aspects of this analysis 
experience significant load imbalance.
There is additional 
interpretation overhead due to the code originally being written in Perl. 
Both issues can lead to analysis of a single
particle taking in excess of 10 minutes. NASA's Large Synoptic Survey
Telescope (LSST) will begin operations in the early to mid 2020s and is
expected to discover and track about 40K Kuiper belt objects over a ten year
survey (compared to the approximately 1000 currently tracked objects).
Thus the need to
do this analysis more efficiently is critical.
In this section, we describe the analysis and its current performance bottleneck

\subsection{Classifying Objects by Their Neptune Resonance}

The objects in the Kuiper belt represent
a record of the dynamical history of the solar system's giant planets. The 
distribution of Kuiper belt objects (KBOs) in orbital resonance
with Neptune is of particular
interest because it can serve as an observational test for theoretical
models of the 
outer solar system's dynamical history \cite{Malhotra1993,Morby2008,Volk2016}.
An object is in orbital resonance with Neptune when there is an integer ratio of 
the number of times it orbits the sun and the the number of times Neptune 
orbits the sun.
When KBOs are observed, their orbits must be analyzed to determine
whether they are resonant. Such dynamical classification is important in
prioritizing objects for continued scientific study. For example,
some specific hypotheses for
the history of the solar system can be tested by observationally
determining the
chemical compositions of resonant KBOs; such observations are costly 
because they can only be done on very large telescopes, and dynamical 
classification is necessary for efficiently planning them.
%%% MMS: below is already said in this section's intro.
%The process by which observed objects are currently being tested for resonant 
%behavior does not scale well to the number of objects that the
%upcoming Large Synoptic Survey Telescope (LSST) is expected to find.
%This paper describes the analysis workload, why the workload does not
%scale, and presents and evaluates alternative implementations.

The classification process, which is detailed in~\cite{Gladman2008}, entails the 
following basic steps:
\begin{enumerate}
  \item The position of the object in the sky at a variety of epochs 
        is determined from observations,
  \item these positions are used to fit an orbit (a combination of position
         and velocity) for the observed object,
  \item the orbit is numerically simulated forward in time under the
         gravitational influences of the solar system's planets,
  \item and the simulated orbital history is analyzed to determine if the object is
        in resonance with Neptune.
\end{enumerate}

If the object is in resonance with Neptune, then the analysis
program determines the specific resonance the object is in
and the amplitude of libration for the associated resonance angle.
%The resonance angle describes the relative positions of the KBO and
%Neptune when the KBO has its closest approach to the Sun. 
The resonance is labeled $p$:$q$ according to the
period ratio between the object and Neptune; a 3:2 ($p=3$, $q=2$) resonance
is one where 
the Kuiper belt object's orbital period is $1.5$ times Neptune's orbital period 
(see Figure~\ref{f:res-example}).
%shows the geometry of such a resonance.
When an object's
evolution is controlled by the resonance, an angle will librate around 
a central value, whereas objects not in resonance will have a resonance
angle that freely circulates between 0 and $360^{\circ}$ as shown in 
the right side of Figure~\ref{f:res-example}. For a given $p$ and $q$, there
are many possible angles determined by the integers $m$, $n$, $r$, and $s$ that could librate;
this is the origin of the nested loops shown in Figure \ref{orbitOverview}. 
%This same analysis can also be applied to theoretical numerical models of the Kuiper belt to produce those
%testable predictions.

%%%%%%%%%%%%%%%%%%%%%%%%%%%%%%%%%

%%%%%%%%%%%%%%%%%%%%%%%%%%%%%%%%%

\subsection{Performance Bottleneck}
The current performance bottleneck for this process is the
last step where the simulated orbit is analyzed for libration of any 
relevant resonance angles. A large number of $p:q$ resonances must be
checked for each object:
all pairings of the integers between $1$ and $pmax$ (on the order of $30$ to $70$), 
where $pmax$ is a runtime parameter, which specifies the granularity of the
angles checked.
%When pmax is higher, a greater number of more granular angles
%are checked for possible resonance.
Objects that are near the correct period
ratio for a large set of $p$, $q$ values take a long time to analyze.

To our knowledge, there is no standard, open source code available to 
performs this analysis.  Researchers generally report
the results of such analyses in papers, but do not make their
codes available or report specific details about the analysis 
methodology. %~\cite{AnalysisResultsCitations}.
We investigate the workload of a Perl script
%that a planetary scientist wrote to perform the analysis.
%% MMS, this submission is not double blind so leave in Kat's name.
that planetary scientist Dr. Kat Volk wrote to perform the analysis.

% NOTE: grouping the small, medium, and long term orbit analyses into one number.
A Serial Perl implementation of this algorithm takes approximately 10 minutes on a 
Xeon Westmere-EP Dual 8-core Processor node
%of the
%University's HPC System 
to fully analyze an 
object that is close to many potential resonance ratios (some of these 
are eventually categorized as non-resonant).
It is not possible to determine up front whether a particle will pass
the initial resonance checks in seconds or require most of the 10 minutes,
thus causing one level of load imbalance in the workload.
Typically, we expect $\sim$25\% of all observed KBOs to be actually
non-resonant and thus require the full analysis time\cite{Petit2011canada}.

\subsection{Future Performance Demands}

In the last 20 years, approximately $1000$ objects have been observed and
classified; because they were typically discovered in
groups of $10$ to $50$~\cite{Bannister2016}, an analysis time
of up to 10 minutes per particle has not been an issue.
However, when the Large Synoptic Survey Telescope (LSST) comes 
online in the mid to late 2020s, the expectation is that forty thousand 
KBOs will be discovered within the first few years of the survey,
and they will be continuously tracked over the survey's ten year 
lifespan~\cite{LSST2009}. Each month the LSST will scan the full 
southern sky multiple times and will provide new
measurements of the position of these objects.  Each set of new 
observations produces a more accurate orbit for the object, 
which requires re-analyzing the orbit for resonant behavior 
(until a sufficiently accurate orbit is determined). 
With 25\% of the 40,000 objects requiring 10 minutes of wall-clock time
each month, the analysis time for one month's worth of data is 
approximately 2 months. This represents a significant performance bottleneck for 
researchers. We also note that code that enables more efficient identification
of
resonant behavior would be useful for analyzing test particles in
numerical simulations of the outer solar system; such simulations 
are used to produce the theoretical predictions the observations 
are meant to test \cite{Levison2008}.
%10000*.60*15: 90000
%#/60: 1500
%#/24: 62.5
%Argument for speed
%\begin{verbatim}
%10000*5: 50000   ! number of objects times number of minutes for a long one. 
%#/60: 833.333333 
%#/24: 34.722222  ! number of days
%would really like to do this on 3 time scales, short, medium, and long term
%might want more precision with more stuff in time files
%\end{verbatim}

%%%%%%%%%%
\begin{figure}
\centering
\includegraphics[]{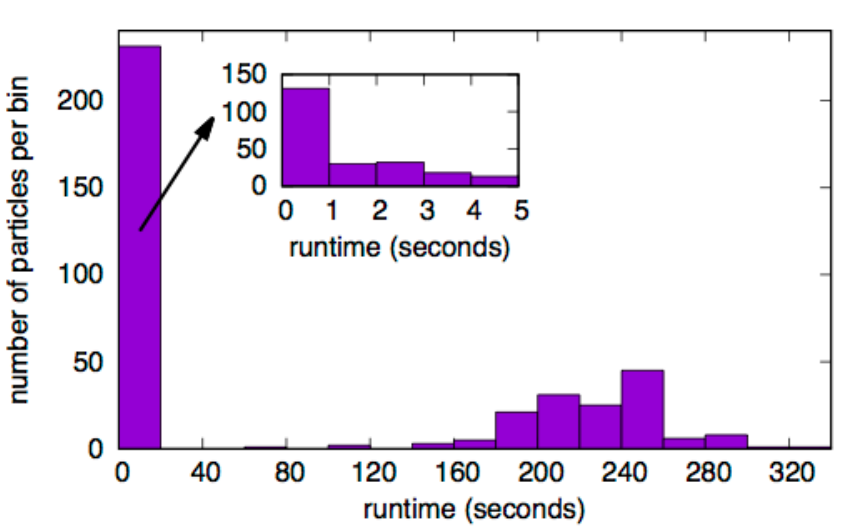}
\caption{Bucketing of runtime for 500 particles. There is a large divide seen between
particles that can be almost immediately rejected, and those that require a significant search time to 
confirm or reject}
\label{fig:perParticleContribution}
\end{figure}
%%%%%%%%%%%%%%%%

%%%%%%%%%%%%%%%%%%%%%%%%%%
\section{Parallelization Problems Due to Irregular Workload}

%% MMS: This can't be removed.  Every section needs an intro
% paragraph.  We need to tell them the punchline before we get
% into details.
The goal of the orbital analysis code is to classify particles as either 
resonant or not resonant with Neptune.
Since there is no communication between the computations for
individual particles, parallelization over the particles would be the simplest.
Figure~\ref{fig:perParticleContribution} shows the per particle contribution 
to execution time on our test set of particles.
The unpredictable analysis times of the particles can easily lead to a small set 
of cores receiving all of the most computationally expensive particles, 
resulting in little to no improvement in the execution time.
Therefore in the next section, we present a more effective strategy for parallelizing
the computation.
This section details our experimental methodology and the load imbalance.

\subsection{Methodology}

We ran experiments on the HPC system of the University of Arizona.
 The machine is an SGI Altix UV 1000 consisting of 170  Nodes, 
 a given node consists of Xeon Westmere-EP Dual 8-core Processors
 running at 2.66 GHz. 
The machine was running Red Hat 6.0 Linux.

The original Perl code and the multithreaded version were run 
using Perl 5.10.1 and using Parallel::Loops from CPAN\cite{petervaldemarmørch2008}.
The OpenMP and Pthreads versions of the code were both compiled 
with g++ (GCC) 4.4.4 20100726 (Red Hat 4.4.4-13).
Chapel was compiled using the Cray Chapel Compiler version 1.12.0.
Python was run using Python 2.7.9, and the 
Multiprocessing module from the Python standard library.

Both of the datasets we used consist of sets of particle simulations. Each particle
has its run discretized into a number of time steps, which records
the relevant angular information at the given point in the simulation.
The more time steps, the finer grained the analysis and the longer processing takes. 
The first, used for the bucketing stage, consists of 500 particles 
whose orbits were divided into 9629 time steps.
This was used only for one step due to the original prohibitively 
long testing time of the 500 particles.
The second, consists of a set of 82 particles, each of 9629 
time steps, this was used for all cross-language execution time comparisons.
The third consists of 100 particles consisting of 50,000 times steps,
analysis of which is currently too time consuming to perform.

\subsection{The Algorithm}

The resonance check for a given particle consists of the following steps as
seen in Figure~\ref{orbitOverview}:
\begin{itemize}
\item %It looks for a constant average semi-major axis. 
      If the {\tt isConsistent()} call on Line 3 returns false the particle can 
      be immediately rejected as non-librating. 
\item Checks all unique ratios of $p:q$ on Lines 8-10.
\item Calculates each possible angular variation of the particle based on 
      the $(p,q,m,n,r,s)$ values   
      created in Lines 8-14 and calls {\tt checkLibration()} to see if the particle
      librates with those angles.
\end{itemize}

As mentioned, the analysis of
each particle is fully independent from that of any other particle, 
and so the particle level was an obvious place for parallelization.
In practice, this consists of parallelizing the {\tt for} loop 
around the call to {\tt CheckRes()} as seen in the pseudocode 
in Figure~\ref{orbitOverview}.
Each particle is checked using the {\tt CheckRes()} function, 
which does all the necessary work to determine if a particle is resonant or not.
%%% MMS: not really necessary.
%This is clean, can be represented in each language with only 
%a few lines of code, and can easily be scaled to allow 
%for testing of a larger number of particles at once.

%\begin{figure}[t]
%\begin{lstlisting}[frame=single]
%parfor each part in particle
% checkResonance(part)
%\end{lstlisting}
%\caption{The naive parallelization of the code. Each particle can be analyzed
%indepently of each other, allowing us to parallelize the outer most loop of the
%code. All of the internals will remain unchanged from original code.}
%\label{NaivePar}
%\end{figure}

\begin{figure}[t]
\centering
\includegraphics[]{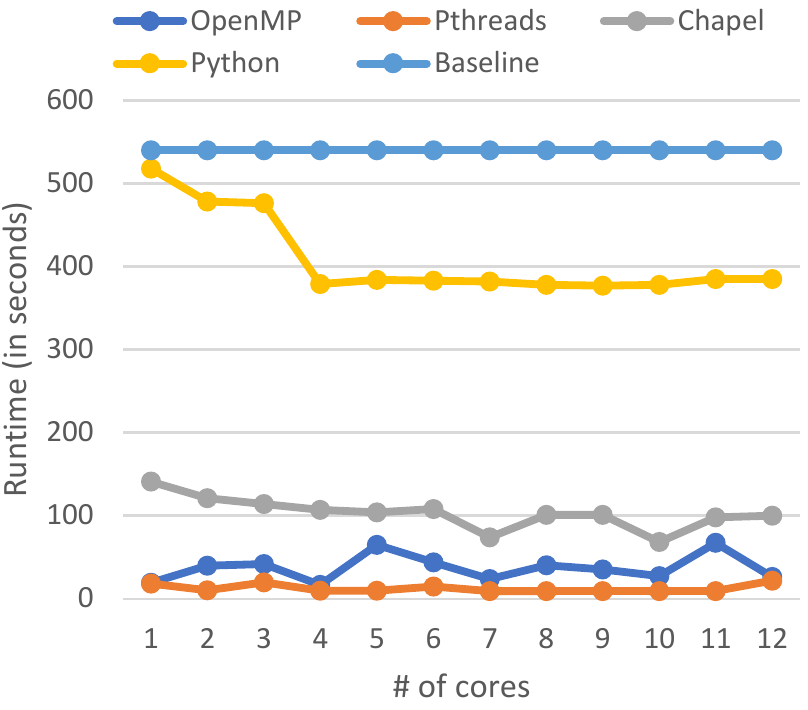}
\caption{Execution times for parallelizing over
the particles.
%for particle level parallelization. 
%These tests consist of dividing up
%the particles amongst processes and having each core 
%handle their set of particles serially. 
Here we see erratic results due to the irregular workload that
results in an unbalanced distribution of the long-running particles.}
\label{fig:particleLevel}
\end{figure}

Unfortunately, this did not improve performance much in the best case, 
and increased execution times in the worst case.
Figure~\ref{fig:particleLevel} shows the behavior of the this 
method in each implementation as the number of threads increases.
The behavior is irregular, with several implementations showing
fluctuations in execution times as the number of cores changed.
%%% MMS: too informal
%This strongly indicated to us that this type of parallelization was inappropriate and that a better understanding of the data was required.

%%%%%%%%%%%%%%
%%%%%%%%%%%%%%%

\subsection{Workload Characterization}

The particles can be placed into three categories:
\begin{enumerate}
\item the range of the particle's semi-major axis is too great 
and therefore can be rejected outright,
\item the particle is in resonance, or
\item the particle cannot be outright rejected, but is not in resonance.
\end{enumerate}

Each of these categories in turn tends to affect execution time in different ways.
The first case is the fastest, contributing a negligible amount of time to overall execution time.
The second case is highly variable, the search space is checked until a resonant angle is found.
Though usually near the beginning of the search, it can theoretically 
require analyzing all possible ratios to find a single resonant angle.
The third is the worst case, a particle that does not resonate but cannot be rejected outright.
The particle at this point requires analyzing the entire search space before it can be rejected conclusively.
As the search space is quite large, these types tend to dominate run time.
All three of these particle types are spread throughout the data and one cannot know which category it belongs to without actually running the code.

The analysis times for individual particles were examined further. 
The original Perl code was used to classify a set of 500 particles, 
and analysis time for each particle was recorded.
Figure \ref{fig:perParticleContribution} shows the execution time 
distribution for analyzing the resonance of these particles.
The results are stark, showing a bimodal distribution of particles 
either finishing nearly instantly, or running for several minutes at a time.
This high variability in the work required to classify a 
particle explains the poor performance of the naive parallelization.
The distribution of the analysis times cannot be known ahead of time,
and so any gain from this type of parallelization will
happen randomly, depending on how the compiler chooses to 
distribute the particles.

%%%%%%%%%%%%%%%%%%%%%%%%%%%%%
\section{Particle-Internal Parallelization}

Parallelizing the computation that occurs within each particle
will help avoid the significant load imbalance issues that occur
between particles.  In this section, we present an approach
to find wavefronts of parallel computation within the analysis
for each particle.

\subsection{Issues with Straight-Forward Parallelization}

The most obvious means to parallelizing the analysis within each 
particle would be to parallelize
work at the outermost $p$ and $q$ loops.
This has two issues due to requirements of the
algorithm.

One issue is early termination.
In the serial code, as soon as a result is found the code returns. 
By doing this the code can often avoid much computation.
As the code searches for a solution, it can often find a valid result
on the first several checks. In the worst case, there can be over 270,000
such checks when $pmax=30$. 
If we cannot return early, we are forcing all particles
to be searched exhaustively. This increase in work leads to worse 
performance, even with parallelism added.

The other issue is one of ordering. As the algorithm progresses it checks 
possible values in a specific order. The earliest lexicographic iteration in 
this space is the least likely to be a false positive for resonance. 
This means that models that allow communication to end early, but cannot
guarantee ordering of results are also invalid for our purposes.

\subsection{Wavefront Parallelization}

The load imbalance and deeply nested structure prevent consistent 
performance gains from being achieved
across the different categories of particles.
To overcome this, the inner most calls were factored out, 
and the nested structure was flattened
into a single list that could be iterated over easily.
The pseudocode for this is given in Figure~\ref{alg:InternalPseudo}. 
The internal parallelism algorithm can be broadly broken into the following steps:
\begin{itemize}
\item{An array of tuples is created, $(p,q,m,n,r,s)$, instead of calling {\tt CheckLibration()} directly,}
\item{CheckLibration() is mapped in parallel over the array of tuples,}
\item{a new array is returned of the results of each CheckLibration() call, and}
\item{these results are scanned serially to find the first occurrence of resonance, or lack thereof.}
\item{If a result is found, the tuple associated with that result is returned.}
\item{Otherwise, we continue searching until all possibilities are exhausted.}
\end{itemize}

This requires extra overhead in storing the tuples,
but it fulfills all of the restrictions discussed.
Returning the results as an array means that it 
does not matter how the individual processors run, the order of the results 
are the same as if it had run serially. We can then perform a serial
scan of the results, finding the one that occurred first lexicographically.
This allows us to reconstruct which result is the optimum in the case of false
positives being returned. 

%%% MMS: don't think we need this
%This method does add one further problem to overcome. If we build up 
%a list of the entire search space at once then every parallel search 
%will have to be over the entire possibility space. 
%This leads to 237,336 possibilities having to be checked independently. 
As previously discussed, as soon as a result is found 
the code can stop. Often libration is found before searching the 
entire space. Building up the entire search space heavily penalizes
those that could have terminated early.
%Over a whole set of particles both of these situations can lead to worse 
%performance than the serial code. 

To solve this a final addition to the algorithm was added. Instead 
of building up the whole search space at once, we build up a 
subset (or wavefront), where some common prefix of the $(p,q,m,n,r,s)$ tuple 
is kept constant. We than check all tuples in this subset in parallel. 
If a result is found we can terminate early, if 
no result is found a new subset can be generated in check. This 
partial generation technique allows for us to parallelize over the 
search space, but without forcing searching the entire space.

%%%%%%%%%%
\begin{figure}[t]
\centering
\includegraphics[]{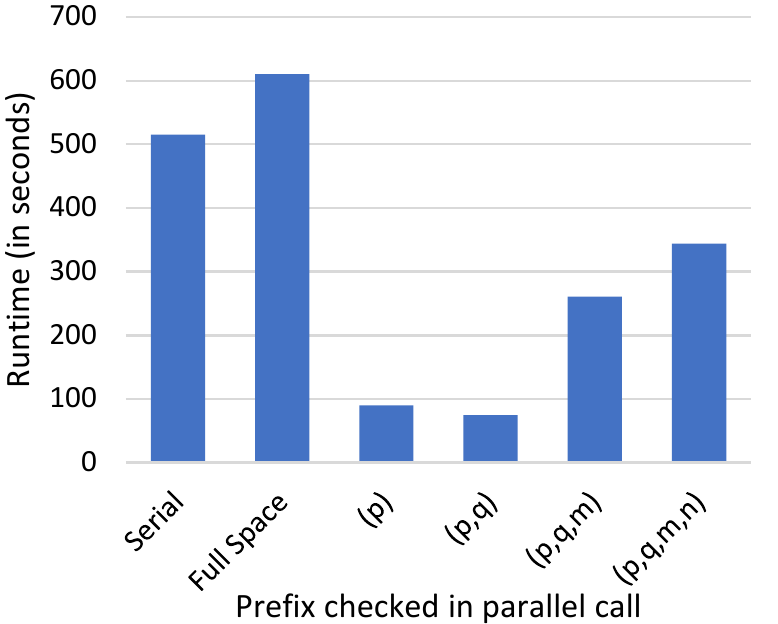}
\caption{Execution times of Python code parallelization when the we parallelize over a subspace. The subspace is built by building a list of tuples. 
Each bar represents building a subspace where the tuple listed is a common prefix in that subspace.
}
\label{fig:LoopTests}
\end{figure}
%%%%%%%%%%

\subsection{Placing the Parallelism}

Determining the correct prefix to keep for each check is difficult to 
determine a priori. If you check on each iteration of $p$
than each check is still checking a very large portion of the search 
space, and so the benefits of being able to return early are lessened. 
If you stop and check on each iteration of $n$ loops you face the opposite 
problem, you end up making so many parallel calls the overhead costs hurts the results. 

An empirical approach was used to best identify where to implement the parallel subset search. 
Figure~\ref{fig:LoopTests} shows the execution times over the 82 particle set. 
This test was done in Python.
% due to ease of refactoring for different experiments.
On the ends we see the expected bimodal behavior, 
the shortest possible common prefix does poorly, as does the longest.
As we slide towards the middle we see significant improvement in overall performance.
The $(p,q)$ prefix outperforms the others, and was chosen as the placement for parallelism
for all further experimentation.

%%% MMS: repetitive
%The algorithm as described allows us to factor out many of the issues related to nested parallelism.
%This opens up the chance for more direct implementation of parallelism, and allows for parallelism
%that is not limited to languages the directly enable nested parallelism.

%%%%%%%%%%%%%%%
%\begin{figure}[t]
%\begin{tikzpicture}
%\scriptsize
%\begin{axis}[
% ybar, 
% ymin=0.0,
% %enlargelimits=0.15,
% %ybar interval=0.5,
% ylabel={Run time (in seconds)}, 
% xlabel={Space checked in each parallel call},
% symbolic x coords={Serial, Full Space, {(p)}, {(p,q)}, {(p,q,m)}, {(p,q,m,n)}}, 
% xtick=data, %nodes near coords align={vertical}, 
% xticklabel style = {font=\small,},
% ytick={0,100,200,300,400,500,600,700},
% xlabel style = {font=\small},
% ylabel style = {font=\small}] 
% 
% \addplot coordinates {(Serial, 515) (Full Space,610) ({(p)},90) ({(p,q)},75) ({(p,q,m)}, 261) ({(p,q,m,n)}, 344)}; 
%\end{axis}
%\end{tikzpicture}
%\caption{execution times of Python code parallelization when the we parallelize over a subspace. The subspace is built by building a list of tuples. 
%Each bar represents building a subspace where the tuple listed is a common prefix in that subspace, with 12 threads.
%}
%\label{fig:LoopTests}
%\end{figure}
%%%%%%%%%%%%%%%

%%%%%%%%%%%%%%%%%%%%%%%%%%
\section{Implementing the Parallelization in Various Programming Models}

The original code written by the planetary scientist was written in Perl.
%The astronomy research community has increasingly been using Python;
%for example, tools such as community maintained Astropy package
%are increasingly being made available \cite{astropy2013}.
In the context of a graduate parallel programming models
course held in the Spring of 2016,
we implement the designed parallelization of the libration analysis code 
in the following
parallel programming models:
\begin{itemize}
\item Perl Thread Library,
\item C++ and OpenMP,
\item C++ and Pthreads,
\item Python Multiprocessing Library, and
\item Chapel.
\end{itemize}

We experimented with a number of parallel programming models with the
primary evaluation metric being the performance.
However, the maintainability and probable evolution of the algorithm
by the planetary scientist is also a consideration, so we recorded
the source lines of code and present code snippets to compare the different
implementations.
% and some observations about how difficult it was
%to implement the parallelization strategy in each language.

\subsection{Parallel Perl}

Due to the algorithm being originally implemented in Perl work 
was done to parallelize in Perl. 
This was problematic though, and Perl was ill-suited to this task.
The multithreading library of Perl is officially recommended 
against by the language designers.
Each new thread spawns an entire new instance of the Perl interpreter, 
a  heavyweight action.
The results of this was that even with the full 12 cores of the system 
the execution time was over 10 times slower than serial implementation. 
The performance and source lines of code count were not recorded for 
Parallel Perl.
%After determining that the parallelization was ill-suited to this problem,
%no further work was attempted within Parallel Perl.

\subsection{C++}

C and C++ are popular languages for writing high performance code. 
%It contains many features for low level optimization, and so has many parallel models implemented over it.
For this comparison two of the most common parallel models were chosen:
\begin{itemize}
    \item OpenMP: A library consisting of a set of compiler directives
    that the user uses to indicate where parallelism will be inserted.
    \item Pthreads: A library for manual thread managements, which 
    enables the user to specify all details about how to divide up 
    information and send it out to different threads for processing.
\end{itemize}

First, the Perl code was translated to equivalent C++ code. 
From this new base code, the internal parallelization algorithm was implemented
using each of the two models.

\begin{figure}[t]
\lstset{morekeywords={parfor},language=C,numbers=left,numberstyle=\small,xleftmargin=5.0ex}
\begin{lstlisting}[frame=single]
for(int p=0;p<pmax;p++)
 for(int q=9;q-p<pmax && q>=0;q--)
  if(checkRatio(p,q,ratios){
   int c = 0;
   for(int m = p-q ; m >= 0 ; m--){ 
    for(int n = p-q-m ; n >= 0 ; n--){
     for(int r = p-q-m-n ; r >= 0 ; r--){
      int s = p-q-m-n-r;
       angles[c]=new Subspace(p,q,m,n,r,s);
       c++;

   bool results[c];
   #pragma omp parallel for shared(result)
   for (int i=0;i<size(angles);i++){
    checkLibration(&angles[i],&results[i]);}

   for(int i=0; i < size(result) ; i++){
    if(result[i])
      return angles[i];}
\end{lstlisting}
\caption{Internal Particle Parallelization using OpenMP}
\label{OpenMPInternal}
\end{figure}

\subsection{OpenMP}

OpenMP is a parallelization library and compiler for C, C++, and 
Fortran \cite{dagum1998openmp}. 
It represents parallelism as a set of compiler directives known as \textit{pragma}.
The user must identify places where parallelism is to be inserted, and then inserts a pragma
that the compiler uses to parallelize the program.

The compiler  then does the necessary conversions to divide the iterations of the loop and assign
it to individual threads, which will then be run in parallel.
In many real world cases, more information is required for the compiler to perform its
work correctly, and so there are a variety of possible arguments and different pragma
that can be indicated by the user. In all cases though, OpenMP is dependent
on the user identifying and specifying the type of parallelism to be used.

Figure \ref{OpenMPInternal} shows the OpenMP implementation of our algorithm.
For our code we first perform the build up of  a list of
tuples, as shown in Figure \ref{alg:InternalPseudo}. 
Because data is being written to a shared memory variable, \textit{result}
we must be cautious of the shared variables since it may lead to 
correctness issues if not handled.
This takes the form of specifying which variables are private to each thread, and thus can be copied,
and which are shared, in which case the compiler must handle writing to a shared memory space.
Finally, since OpenMP does not allow branching in or out of a structured block, we place the 
results in a global list of results. This list is scanned to see if resonance was found, 
and then to identify which values led to resonance.

%%% MMS: think we will 
%The OpenMP style of directive-based parallelism is simple and clean.
%It allows the user to specify parallelism where it exists in their current code,
%without forcing the user to refactor their code around the parallelism. 
%For programmers already working in C, C++, or Fortran, it presents a viable model
%for easy parallelization.
%As more programmers transition to using high-level, highly productive languages, such as Python and Perl,
%this option becomes less appealing. 
%Unless the performance is absolutely needed, users will tend to shy away from converting to a compatible language.
%This necessitates finding other parallel models that align with trends in programming language usage.

\begin{figure}[t]
\lstset{morekeywords={parfor},language=C,numbers=left,numberstyle=\small,xleftmargin=5.0ex}
\begin{lstlisting}[frame=single]
// ti is the array of thread ids
pthread_t* ti = new pthread_t[threads];

for(p=1;p<pmax;p++)
 for(int q=p;p-1<pmax && q>0;q--)

  // inspecting for tuples of work
  if(checkRatio(p,q,ratios))
   for(int m=p-q;m>=0;m--)
    for(int n=p-q-m;n>= 0;n--)
     for(int r=p-q-m-n;r>=0;r--)
     {
      int s=p-q-m-n-r;
      tuple.p=p;tuple.q=q;tuple.m=m;
      tuple.n=n;tuple.r=r;tuple.s=s;
      subset.push_back(tuple);
     }
   
   // spawn threads for parallelization
   threads=min(subset.size(),max_threads);
   for(int i=0;i<threads;i++)
   {
    err=pthread_create(ti+i, NULL,
    partialCheck,(void*)(&i));
   }
   for(int i = 0 ; i < nTh ; i++)
    pthread_join(ti[i],NULL);

   // gather the results
   for(int i=0;i<subset.size();i++)
     if(results[i])
       return subset[i];
\end{lstlisting}
\caption{Internal implementation using Pthreads. This model requires all 
details to be managed directly by the programmer, and thus requires the most 
additional code out of any of the programming models discussed. }
\label{pthreadsspjo}
\end{figure}

%%%%%%%%%%%%%
\subsection{Pthreads}

Pthreads is a low level C/C++ library for creating shared memory multi-threaded programs.
%The library provides few helpers in any aspects of data parallelism.
%%There are few helpers in Pthreads for 
%%automatically partitioning the data or the computation space of a problem. 
Programmers define and create the number of execution threads, 
partition the computation and the data, and explicitly define 
which thread is going to do what part of the computation. 

Figure~\ref{pthreadsspjo} shows the internal parallelization using Pthreads. 
Just like other models, we create a subset of computation space
that we want to check. Next we need to create a number of 
threads that are going to process part of this subset. In Pthreads model each
thread's 
execution starts from a function that we pass in to \texttt{pthread\_create()} constructor.
%%% MMS, not sure what was going on here.
%here called partialCheck. The partialCheck function contains the code 
%from the CheckLibration function from original serial code, with minor changes
%to perform the explicit data partitioning required by the model.

%%% MMS: We are going to leave this out and just include the SLOC graph.
%Implementing the proposed parallelization algorithm for our particular 
%code was fairly straight forward. There were no need for 
%data communication among threads, once the tuples are built the problem
%becomes embarrassingly parallel.
%However parallelizing code using Pthreads is not a trivial 
%task in general. There are two aspects to Pthreads that allow for high performance: 
%first, it is a low level library,
%second, it is high flexible since everything can be manipulated by hand. 
%However, the same aspects makes it very hard to program with Pthreads, 
%forcing the programmer to handle all details of the problem.
%This leads to it being less productive than the other parallel programming models 
%used in this paper.

%\begin{figure}[t]
%\begin{lstlisting}[frame=single]
%int range = subset.size();
%int thread_job = range / threads;
%int start = th_job * id;
%int end = min( (start + thread_job) ,range );
 
%for(int i = start ; i < end ; i++)
%{
%   checkLibration(
%}
%\end{lstlisting}
%\caption{Partial code inside each thread function (partialCheck), for dividing the computation.}
%\label{pthreadsdatapart}
%\end{figure}

\begin{figure}[t]
\centering
\begin{lstlisting}[frame=single,language={Python},basicstyle={\small\ttfamily},numbers=left,numberstyle=\small,xleftmargin=5.0ex]
for p in range(0,pmax+1):
 for q in range(p,0,-1):
  subset = []
  if( (p,q) not in ratios):
   for m in range(p-q, -1, -1):
    for n in range(p-q-m,-1,-1):
     for r in range(p-q-m-n, -1, -1):
      s = p-q-m-n-r
      subset.append((p,q,m,n,r,s))
   pool = MultiProcessing.Pool()
   pf = Partial(CheckLibration, part)
   results = pool.map(pf, subset)
   for i in range(len(results)):
    if results[i]:
     return subset[i]
\end{lstlisting}
\caption{Particle Internal Parallelization in Python. The usage of a map for parallelization
requires modifying the code to a fit map semantics. In our case, because we have a collection
of tuples, but a fixed particle for each set, we are forced to use a partial
function fit into the map model. }
\label{PythonInternal}
\end{figure}

\subsection{Python}

Python has become one of the most popular languages for 
planetary scientists~\cite{momcheva2015software}.
Due to the usage of a Global Interpreter Lock (GIL) in the reference interpreter
for the language parallel options are limited.
The GIL prevents multiple threads from executing Python bytecode at the same time.
The Multiprocessing library in Python sidesteps the GIL by using subprocesses rather than threads.
This is more heavyweight than threading in other languages, but is a necessity forced upon
the users by the GIL.

Figure~\ref{PythonInternal} shows the snippet for the particle-internal
 parallelization in Python.
In the Python Multiprocessing library, the standard form of data parallelism 
is represented as a map function called over a collection. 
%This can be unwieldy for users 
%unfamiliar with map semantics.
%This is in line with the functional elements that Python has
%increasingly chosen to pursue.
In this case, the {\tt CheckLibration()} function will be mapped over the subset of tuples 
the inspector has collected.
The results are guaranteed to be ordered the same as the input tuples' orders, allowing 
us to guarantee our search for the optimum.

There is a mismatch between the angles that change from each call, and the particle that remains fixed. 
This led to us having to create a partial function to combine the checkLibration() function and the particle data. 
This creates a new function which has the same execution as checkLibration, 
but with the value for particle held fixed across all calls.
%This is not an obvious construct for many users who are unfamiliar with functional programming concepts.
%Having to handle this mismatch between collections and consistent values across the map make parallelizing the code more difficult.

%%% Leaving out and will just talk SLOC later.
%Thus, we see an imbalance in the language.
%Serial Python is a highly productive language, and for this reason has seen widespread adoption in scientific computing.
%The astronomy research community has increasingly been using Python;
%for example, tools such as community maintained Astropy package
%are increasingly being made available \cite{astropy2013}.
%The conversion to a parallel model is not trivial, 
%can require significant refactoring in certain cases, 
%and can require working with new and exotic seeming functional constructs. 
%This can often dissuade users from converting their serial Python
%code to parallel.
%For users already committed to using the language though, it presents
%a possible means to gain significant speed-up in CPU bounded
%problems. 
%Finding a means to ease users across the gap from serial to parallel Python 
%could enable more programmers to utilize these features.

\begin{figure}[t]
\centering
\lstset{morekeywords={forall},numbers=left,numberstyle=\small,xleftmargin=5.0ex}
\begin{lstlisting}[frame=single]
var subset:[0..size];
for p in [0...pmax]{
 for q in [p...0]{
  var counter : int = 0;
  if( checkRatios(p,q,ratios) ){
   for(m in [(p-q)..0] by -1){
    for(n in [(p-q-m)..0] by -1){
     for(r in [(p-q-m-n)..0] by -1){
       var s : int = p-q-m-n-r;
       subset[counter] = (p,q,m,n,r,s);
       counter += 1;
       size += 1}}}}}}
   forall( i in 0..#counter )
    found[i] = CheckLibration(subset[i]);
   for( i in 0..counter){
    if( found[i] )
      return subset[i];}
\end{lstlisting}
\caption{The final Chapel implementation. The single indicator of parallelism in the code
is the usage of the \textbf{forall} keyword instead of a \textbf{for} keyword.}
\label{ChapelCode}
\end{figure}

\subsection{Chapel}

Chapel is the newest programming language of the three, a model designed
from the ground-up to support parallelism~\cite{ChapelOverviewJan13}. 
Created by the super-computer
manufacturer Cray, Chapel represents an attempt to integrate parallelism from the ground up into a language. 
Parallel features in this language are syntactic elements of the language,
with keywords indicating when parallelism is to be used.
%This syntactic integration ease usage of parallelism,
%and help with some of the mental blocks that lead
%users to underutilize parallelism. 

The language allows us to match closely to the pseudocode algorithm
originally show. We build a set, we have a parfor over that set, 
and then we check the results for a positive result.
The difference between the serial version, and the parallel version 
is as simple as replacing \texttt{for} with \texttt{forall}.

\subsection{Source Lines of Code}

%%%%%%%%%%
%\begin{figure}[t]
%\includegraphics[]{images/SLOC_cropped}
%\caption{Source Lines of Code for each Language. These were developed across three programmers and
%%so direct comparison of languages is difficult. The data is shown to represent an approximation of each implementations complexity.
%}
%\label{slocGraph}
%\end{figure}
%%%%%%%%%%%

%To approximate the difficulty of working within each parallel model 
We 
also compare the total Source Lines of Code (SLOC) for the serial and parallel
versions of each language.

%On average, the longer a code is the more difficult it becomes for the programmer
%to reason over its functionality, as well as forcing the programmer to write more. 
%This all adds to the overall time required to write and maintain a codebase, and so there is an advantage in languages that allow more compact representation of 
%execution.
%All data for our SLOC comparison was generated using
%SLOCCount\cite{sloccount}. 

Python requires the least amount of code, but required certain constructs, such as partial functions,
which may be unfamiliar to users from a imperative or object oriented background.
This is a case where the difference in SLOC can hide the actual complexity 
of the implementation from
readers.

Though Chapel is more verbose than Python, it still has fewer lines of code than the serial Perl.
The transition to the parallel algorithm was also simple, with the final 
parallelization change requiring only a keyword change from {\tt for} to {\tt forall}. 
Both the Python and Chapel code also had smaller differences between 
the serial and parallel versions
than with the C++ versions.

For the OpenMP version the majority of the difference came from the refactoring necessary 
to fit the final parallel algorithm chosen for this task. 
The actual insertion the code related to OpenMP was much less: 3 lines overall including
the pragma, the import, and a required initialization statement to control number of cores used.

PThreads had the largest difference by a wide margin. 
Beyond the changes required for the parallel algorithm, significant changes were 
required to fit that within the framework of PThreads.
Explicitly managing all of the threading logic leads to significant tangling
with the underlying logic of the chosen algorithm.

All 4 versions required some amount of refactoring, both to implement the new algorithm and to
add the parallel logic.
The Chapel and OpenMP versions were the least invasive in the insertion of parallel logic. 
And only these two allow for switching between parallel and serial code without requiring 
any changes to the final code. 
In this regard, they have the much less tangling than the other two implementations.

%boilerplate that comes with that, it also
%required a less verbose implementation than Perl, coming in below the baseline.
%The parallel Perl version was in line with both the Chapel and Pyth%on versions. 
%All three of required comparatively little refactoring and additions compared to the
%implementations on top of the C++ versions.

%The OpenMP model was implemented on top of C++ code, and required little overhead to implement.
%OpenMP consists largely of Pragma statements and refactoring of the code
%to make those Pragma fit appropriately with the parallel model, 
%because of the usage of compiler pragma rather than explicit
%library calls the onerous is pushed from the user to the compiler.
%The most extreme was Pthreads, which requires explicit handling of most of the aspects of the multithreaded code. 
%This level of detail conveys power to the user, but also puts all 
%responsibility upon the user, it also means that quickly changing 
%aspects of the parallel model
%can become quite arduous as large sections of code must be 
%refactored anytime the model is to be changed.
%%%%%%%%%%%%%%%
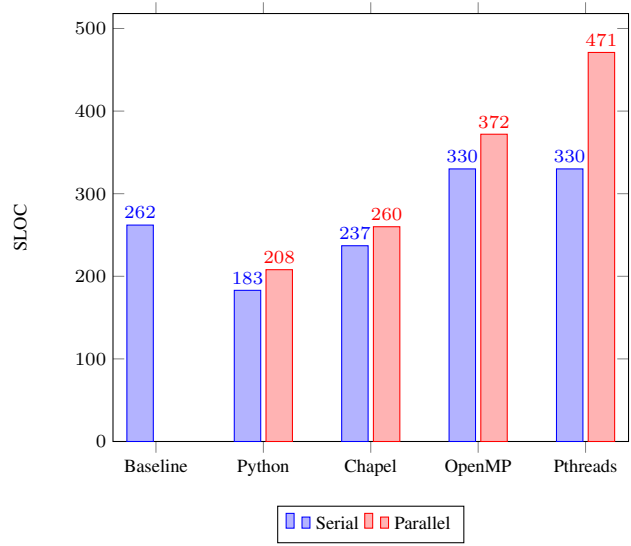
\begin{figure}[t]
\begin{tikzpicture}
\scriptsize
\begin{axis}[
 ybar, 
 ymin=0.0,
 legend style={at={(0.5,-0.15)}, 
 anchor=north,legend columns=-1}, 
 ylabel={SLOC}, 
 symbolic x coords={Baseline, Python, Chapel, OpenMP, Pthreads}, 
 xtick=data, nodes near coords, nodes near coords align={vertical}, ] 
 
 \addplot coordinates {(Baseline, 262)  (Python,183) (Chapel,237) (OpenMP, 330) (Pthreads, 330)}; 
 \addplot coordinates { (Python,208) (Chapel,260) (OpenMP, 372) (Pthreads, 471)};
\legend{Serial,Parallel}
\end{axis}
\end{tikzpicture}
\caption{Source Lines of Code for each Language. These were developed across three programmers and
so direct comparison of languages is difficult. The data is shown to represent an approximation of each implementations complexity.}
\label{slocGraph}
\end{figure}
%%%%%%%%%%%%%%%

%%%%%%%%%%%%%%%%%%%%%%%%%%
\section{Performance Comparison}

We tested the implementations on two different datasets.
The first was the same test set used for the Perl baseline data. 
This consists of 82 particles, represented as location and orientation at 9629
different time steps in the simulation.
The long run-time at this granularity prevented analysis of the particles at a finer level
of time.
Each version showed at least 6x speed-up over the baseline code,
with the OpenMP version achieving 103x speed-up, and a final runtime of 5.2 seconds versus the baseline 
run of 541 seconds.

Due to the success in improving performance on the dataset we did further testing
on a second dataset of particles.
These modeled each particle at 50,000 time steps and represented the granularity that 
the scientists would prefer to analyze.
With the baseline Perl version a single particle could take more than 45 minutes to analyze.
Performing analysis at this level was even less feasible for the incoming amount of data.
For these long running particles we were able to get significant improvement as well.
In the worst case, the Python version had 8.4x speed-up.
In the best case, the OpenMP version once again performed best and gave us a 105x speed-up, bringing
the worst particle down from 45 minutes to 25 seconds.

Finally, we give a Source Lines of Code (SLOC) comparison across language. 
This is done to give an approximation of the effort required to implement each version
of the code. 
With this metric, both the Python and Chapel version outperformed the baseline Perl code.

\subsection{Check-Internal Parallelization}

The final version chosen for experimentation
uses the Internal Parallelization model over 
the Naive model, because that version was able to handle the irregular
workload.
In this the checks over the search space for a given particle was parallelized, 
this has us handling each particle one by one, but handling
each of them much quicker. These implementations were tested 
on the set of 82 short-length particles.

\begin{figure}[t]
\centering
\includegraphics[]{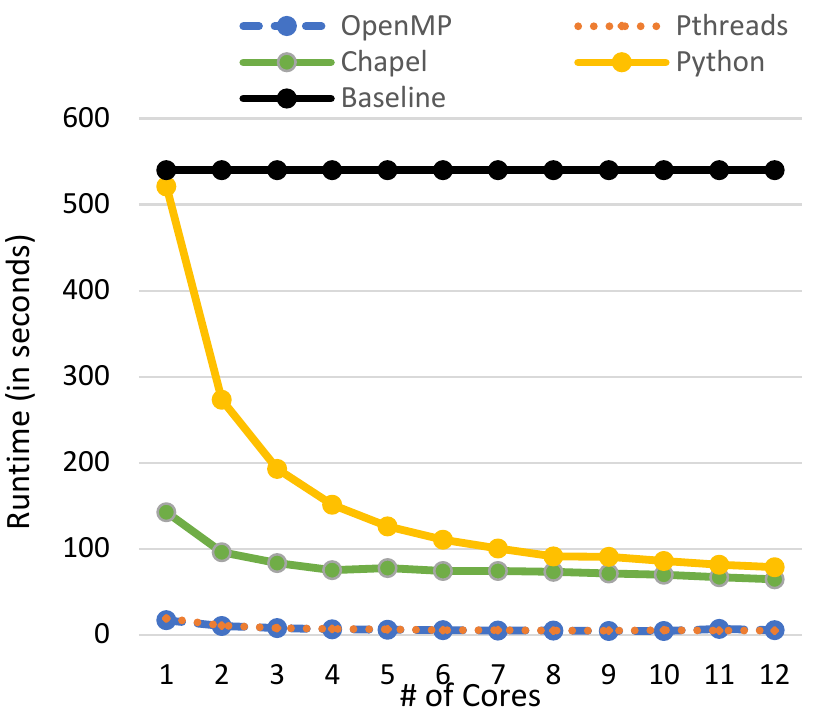}
\caption{Execution times for parallelization of the internal particle check. In this model
each particle is run one-by-one, but search for each particle is parallelized. The Pthreads
and OpenMP versions overlap almost exactly.}
\label{fig:deepparallel}
\end{figure}

\begin{figure}[t]
\centering
\includegraphics[]{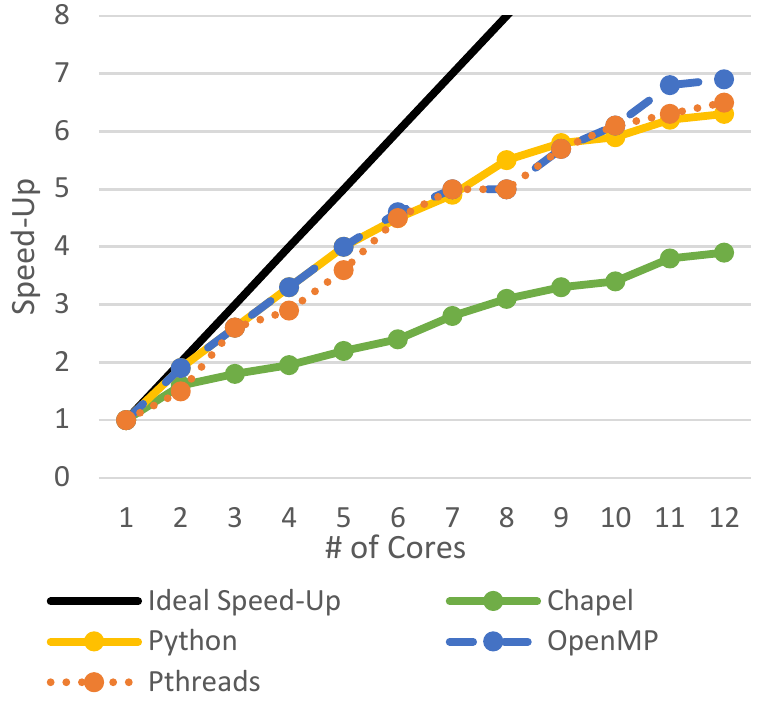}
\caption{Graph showing the speed-up for each implementation. Each implementation is compared to its own serial implementation.}
\label{fig:SpeedUpCropped}
\end{figure}

Figure~\ref{fig:deepparallel} shows the raw execution time changes for each implementation. 
Figure~\ref{fig:SpeedUpCropped} shows the Speed-Up curves for each of the implementations.
With this we are able to handle the unpredictable workload much better. 
The new implementation is entirely lacking in the sharp changes in performance
found in the particle level parallelism. 
Each programming model shows a similar curved improvement as the number of cores increased. This shows that the algorithm chosen improves execution time
of the longest running particles without too heavily penalizing 
those that can quickly return an answer. 
Thus, the flattened tuple structure chosen overcomes the 
workload imbalance issues inherent to the dataset.

The Python version performs well for an interpreted language. 
As the number of cores increases we see that the
 execution time approaches the speed of the compiled Chapel language.
This alone represents significant speed-up, enough to bring the 
total execution time into the realm of feasibility.

The Chapel version performs well with a low number of cores, 
but does not scale as well as other versions. Only barely outperforming
the Python version, surprising considering that Chapel is a compiled language.
Though it did not perform as well as one would hope, the language 
is still young and the team has specifically targeted performance
as their major focus for improvement in the coming years.

Both of the C++ versions match each others performance. 
OpenMP and Pthreads each show the same curve, and in 
general seem to be functionally identical to one another.
Overall these two versions perform the best, with the Pthreads 
version barely out-performing the OpenMP code. 
The final implementation brings then execution time down to 
5.2 seconds from the 541 second baseline serial Perl execution, 
giving a final speed up of 103x.

\subsection{Large Time Scale Particles}

%As previously stated there were two datasets, the first 
%consisting of smaller time-scale particles representative of the
%type of particle currently being analyzed.
Testing was extended to the
second set of longer time scale particles. 
These particles represent a level of detail that the 
researchers would prefer for their analysis, but were unable because of the 
prohibitive execution times.
With the original Perl implementation each particle was 
taking more than 10 minutes to analyze on the low end. On the high end a single particle could took up to 45 minutes.
This quickly led to an unworkable situation.

\begin{figure}[t]
\centering
\includegraphics[]{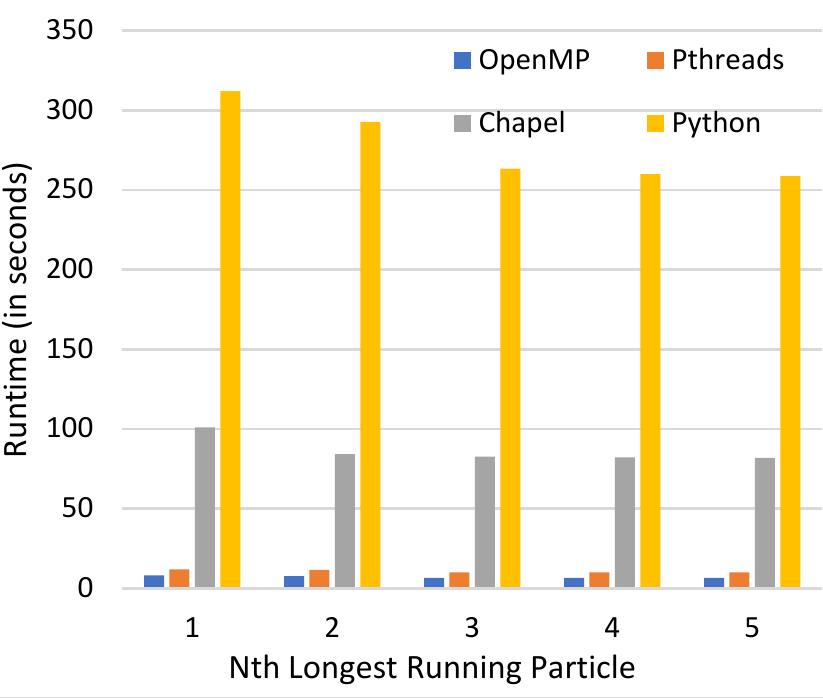}
\caption{Comparison of the execution times of the top 10 longest running particles. Each of these particles are from the set of 100
long-time scale particles. In the original baseline each of these particles takes over 45 minutes on average to analyze. 
Each of our parallel versions bringing the execution time down significantly.}
\label{fig:worstcase}
\end{figure}

Figure \ref{fig:worstcase} shows a comparison of the execution times of 
the 5 worst particles from the long-time scale data set.
The three versions shown perform well, bringing the execution times 
down much closer to the realm of feasibility. The OpenMP version
is the fastest. but even in the worst case there is 8.2x speed-up over 
the 45 minute long per particle time for the baseline serial Perl implementation.
Even with the slowest version we can show that the run-time
can be brought down into a more feasible realm.

This problem of handling a larger dataset is interesting in that it 
is a common problem seen in the Scientific Computing community. 
So often are they limited by the speed at which they can do computation 
that they often work with simplified workloads to bring execution time
down into the realm of feasibility.
And so speed-up for them often represents a means to continue on 
to more complex computation, or finer grained analyses of their data.
In this case, execution time was able to be improved so significantly 
that it brings nor only the current time scale level into a more feasible time frame,
but to allow for better analysis to be done in the future.

 %%%%%%%%%%%%%%%%%%%%%%%%%%
\section{Related Work}

There has been much work on improving the performance of nested
parallelism.  Blikberg and S{\o}revik~\cite{Blikberg2005} 
argued for flattening nested parallelism into a single level of
parallelism.  They then have an approach for load balancing when
there is some model of how much work each task is doing.  In
the orbital analysis application the amount of work that will be
needed per particle cannot be modeled ahead of time.
Thibault et al.~\cite{Thibault2008}  are dealing with the problem 
of expressing data affinity to the underlying runtime system.
The orbital analysis program has severe load imbalance issues,
not data locality issues.
Dimakopoulos et al.~\cite{Dimakopoulos2008} created a 
nested parallelism benchmark
and then showed that many nested parallelism implementations of OpenMP have a 
lot of overhead.

A significant amount of research investigates the advantages and
disadvantages of various programming languages in the context
of scientific computing. 
Caie et al.~\cite{Cai2005} describe various libraries and capabilities in
Python such as NumPY and the ease of calling Fortran and C and how
those impact the performance of stencil computations that occur when solving
partial differential equation solvers.
Many others have compared various parallel programming languages
in terms of their performance and programmability 
with various 
benchmarks and applications~\cite{Mattson03,Sterling04,Cantonnet04,Chamberlain07,Shet08,Karlin13}.
This study focuses on characterizing the workload and performance
alternatives for implementing a specific analysis needed for the
Large Synoptic Survey Telescope (LSST) project.

Shen et al.~\cite{Shen12} describe a parallel algorithm implemented in MPI 
for the analysis of objects that are close to the earth.
The algorithms in question are different than those we study in
this paper. 
%Additionally, we consider the feasibility of the scientist
%being able to quickly prototype in any language as an evaluation
%metric.

%%%%%%%%%%%%%%%%%%%%%%%%%%
\section{Conclusion}

In this paper, we discuss problems that arise for the analysis of
orbital particle data.
We provide analysis of the deeply nested structure, and the sparseness of the search space,
which lead to significant load imbalance in naive attempts at parallelization.
We describe a solution to these issues, presenting an algorithm
for flattening the nested structure to allow better parallelization.

We then discuss the issues that arise during parallelization in Perl, Python, OpenMP, PThreads, and Chapel.
We include code snippets which show the implementation details 
for our algorithm, and the language specific differences that arise because of this.

In the end, we show significant speed-up in all languages, excepting Perl.
Special attention was given to improvements in Python, a language which has
seen widespread adoption in scientific computing.
We achieved 9.1x speed-up in Python, a language whose serial version performed comparably to the baseline.
In the best case we show the Pthreads version achieved a 103x speed-up 
over the original performance.
These improvements allow for analysis of finer-grained data that had previously
been considered infeasible due to the execution time bottleneck.

%We do a performance analysis on an algorithm that analyzes
%the orbits of objects in the Kuiper belt.
%We then present an approach to effectively parallelize the algorithm even
%though it is not possible to determine a priori how to evenly distribute the load.
%We implement this parallelization approach in a number of parallel 
%programming models and determine that .

%The amount of astronomical data available is increasing ever faster. This is creating richer sets of information for scientific analysis, allowing for richer understanding of our universe.
%But it also threatens to overwhelm us, providing us with so much data that we cannot conceivably sift through it all.
%If the data is too large to be worked with than it is little more than noise to researchers.
%Identifying the tools and techniques that allow scientists to efficiently handle the data workloads facing them is of paramount importance moving forward.

%%%%%%%%%%%%%%%%%%%%%%%%%%%
%\section*{Acknowledgment}
%[No idea as well]

%%%%%%%%%%%%%%%%%%%%%%%%%%
\bibliographystyle{IEEEtran}
\bibliography{libration}

\end{document}